\documentclass[%
 reprint,
 amsmath,amssymb,
 aps,
]{revtex4-2}
\usepackage{amsmath}
\usepackage{graphicx}
\usepackage{dcolumn}
\usepackage{bm}
\usepackage{footnote}
\usepackage{subfigure}%

\begin{document}


\title{Acoustic Black Holes in a Shock-Wave Exciton-Polariton Condensate}

\author{Junhui Cao$^{1}$}%
\email{tsao.c@mipt.ru}
\author{Jinling Wang$^{2}$}
\author{Kirill Bazarov$^{1}$}%
\author{Chenqi Jin$^{2}$}
\author{Huijun Li$^{2,3}$}
\email{hjli@zjnu.cn}
\author{Anton Nalitov$^{1}$}%
\author{Alexey Kavokin$^{1,4}$}%
\email{a.kavokin@westlake.edu.cn}
\affiliation{$^{1}$Abrikosov Center for Theoretical Physics, Moscow Center for Advanced Studies, Moscow 141701, Russia.
\\$^{2}$Institute of Nonlinear Physics and Department of Physics, Zhejiang Normal University, Jinhua, 321004 Zhejiang, China\\
$^{3}$Zhejiang Institute of Photoelectronics, Jinhua, Zhejiang 321004, China\\
$^{4}$Department of Physics, St. Petersburg State University, University Embankment, 7/9, St. Petersburg, 199034, Russia}



\begin{abstract}
We demonstrate the spontaneous formation of acoustic black holes in exciton–polariton condensates triggered by discontinuous Riemann-type initial conditions. Starting from a quasi-conservative Gross–Pitaevskii model, we show that nonlinear dispersive shock waves naturally generate spatial regions where the local flow velocity exceeds the speed of sound, creating a self-induced transonic interface that functions as an acoustic horizon. Unlike previous schemes relying on externally engineered potentials or pump–loss landscapes, our approach reveals that the intrinsic nonlinear hydrodynamics of polariton fluids alone can lead to horizon formation. Using Whitham modulation theory and numerical simulations, we characterize the transition between subsonic and supersonic regimes and estimate the corresponding surface gravity and Hawking temperature. This mechanism opens a new route toward realizing polariton black holes and studying analogue gravitational effects, including Hawking-like emission, in Bose-Einstein quantum liquids.
           
\end{abstract}

\maketitle

\footnotetext[1]{$^\dagger$These authors contributed equally to this work.}

\paragraph*{Introduction ---}
The study of exciton–polariton condensates has matured into a rich arena where quantum many-body physics, nonlinear optics, and non-equilibrium statistical mechanics intersect \cite{pieczarka2020,carusotto2013,bloch2022}. Polaritons, hybrid light–matter quasiparticles formed in semiconductor microcavities by strong coupling between cavity photons and excitons, combine light effective masses with repulsive interactions inherited from their excitonic component \cite{kasprzak2006,deng2002,zhang2020}. This combination enables macroscopic coherent phenomena such as Bose-Einstein condensation, superfluid flow, supersolid stripe pattern, to be realized at elevated temperatures and with a high degree of experimental control \cite{zhang2022,byrnes2014,amo2009,amelio2020,lerario2017,trypogeorgos2025}. This mixture of coherence and interactions produces a variety of nonlinear hydrodynamic phenomena, including solitary waves, quantized vortices, and dispersive shock waves (DSWs) \cite{xue2014,bloch2022,lagoudakis2011}.

Using condensed-matter systems to simulate aspects of field theory in curved spacetimes, analogue gravity has emerged as an influential interdisciplinary program. A central concept in this field is the acoustic horizon: in an inhomogeneous flow, the local transition between subsonic and supersonic regimes defines a one-way interface for low-energy sound-like excitations that closely resembles the event horizon of a black hole \cite{unruh1981,solnyshkov2019,garay2000,macher2009,visser1998}. In analogue systems, horizon formation allows probing of phenomena analogous to Hawking radiation using accessible laboratory observables such as correlation functions and emission spectra \cite{weinfurtner2011,drori2019}. Experiments in ultracold atomic gases, nonlinear optics, and open quantum fluids have pursued various realizations of horizons \cite{steinhauer2016,carusotto2013}; in polariton condensates, several experimental proposals and demonstrations have sought to engineer horizons via spatially varying pumps, moving defects, or shaped potentials \cite{jacquet20201,nguyen2015}.

Despite these advances, many schemes to realize horizons rely on external engineering, such as precisely structured pumping profiles, tailored potentials, or moving obstacles, which can complicate interpretation by introducing extrinsic dissipation, disorder, or parameter fine-tuning \cite{gerace2012a,jacquet20201}. An alternative and conceptually appealing strategy is to exploit intrinsic nonlinear hydrodynamics to generate transonic flows autonomously. DSWs, the nonlinear wavetrains that arise when nonlinear steepening competes with dispersion, are a paradigmatic manifestation of such hydrodynamics \cite{el2016d,wan2007d,hoefer2006d}. In conservative superfluids and nonlinear dispersive media, Riemann-type initial-value problems with abrupt jumps in density or phase typically evolve into a structured pattern consisting of a DSW on one side and a rarefaction wave or simple wave on the other \cite{el1995decay}. Crucially, these self-organized patterns can carry regions where the local flow velocity exceeds the local sound speed, thereby creating transonic interfaces without explicitly imposed external landscapes.

In this work, we propose and analyze a minimal mechanism for creating polariton acoustic horizons using Riemann-type discontinuities that trigger DSW formation. Focusing on the quasi-conservative regime which is appropriate for resonant injection and when the exciton-polariton condensate decays on timescales long compared with the coherent-field dynamics, we adopt an effective Gross–Pitaevskii description to capture the essential nonlinear wave physics while retaining analytical tractability. Within this framework, we use Whitham modulation theory to obtain semi-analytic descriptions of the emergent DSWs \cite{kamchatnov1994,liu2022e}, and we carry out time-dependent numerical simulations to validate and extend the analytic picture. We additionally perform Bogoliubov spectral analysis on the evolving background to identify propagating modes and scattering channels, and we compute two-point density correlations (via truncated Wigner approximation method) to characterize pair production consistent with analogue Hawking processes. Our findings reveal that the pursuit of analogue gravitational phenomena and the fundamental study of nonlinear hydrodynamics in quantum fluids are deeply interconnected, with shock waves serving as a natural and potent tool for sculpting the effective spacetime metric of sound.

\paragraph*{Shock-wave in polariton condensate ---}\label{II}

\begin{figure}
    \centering
    \includegraphics[width=0.9\linewidth]{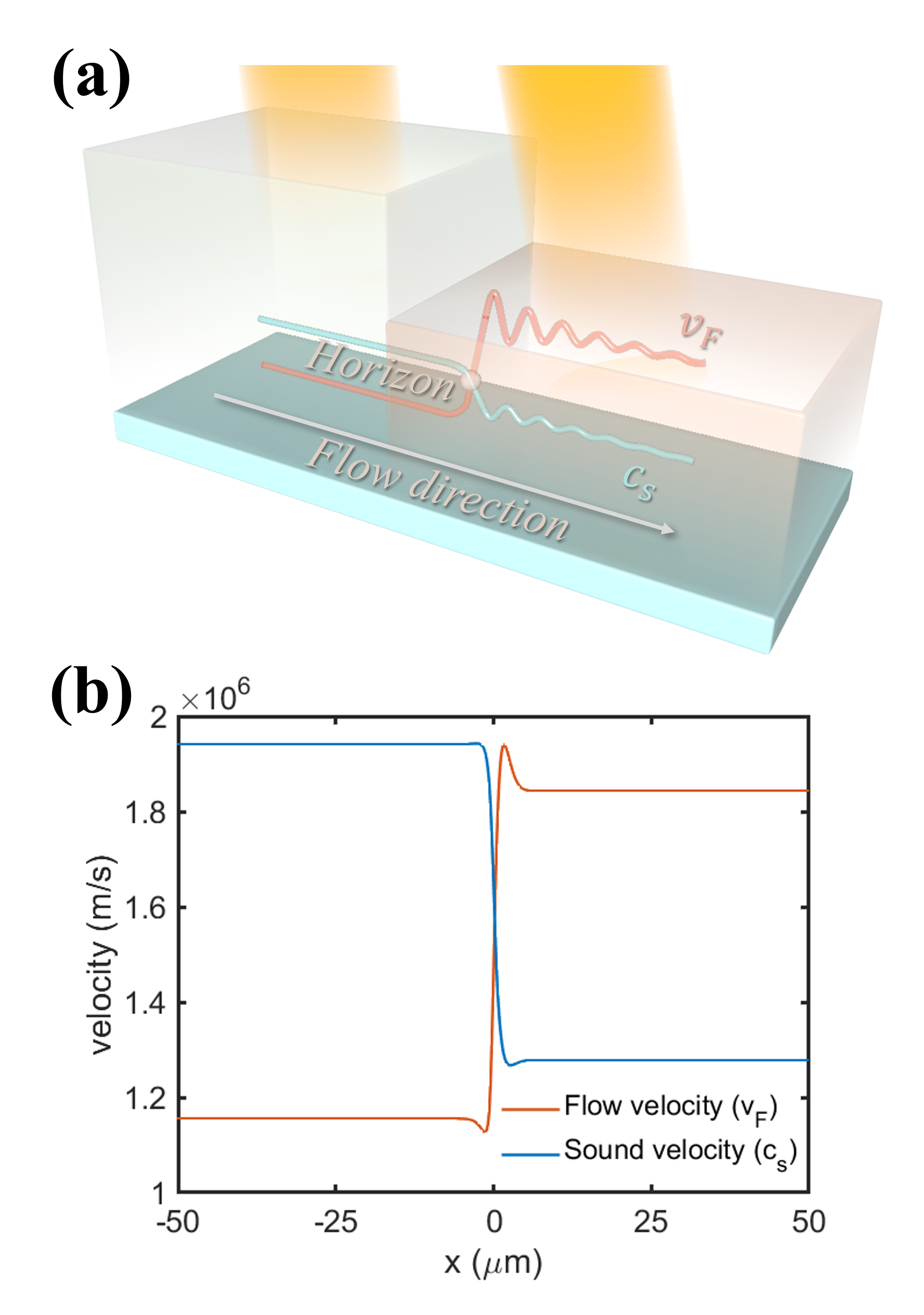}
    \caption{(a) Schematic of the polariton condensate under two resonant oblique pumping beams. (b) Flow ($v_F$) and sound velocity ($c_s$) profile. The acoustic horizon is created at the transonic point $x=0\ \rm \mu m$. $v_F$ and $c_s$ can be precisely controlled by the amplitude and incidence angle of the resonant pump, respectively.}
    \label{fig:scheme}
\end{figure}


We consider a semiconductor microcavity with a sufficiently high quality factor under resonant optical pumping by femtosecond laser pulses\cite{jia2025femtosecond}. The initial state of the system is prepared by a spatially structured double laser beam that imposes a step-like profile in the polariton density and flow velocity, corresponding to the Riemann-type initial conditions described below (Fig.~\ref{fig:scheme}). Exciton-polaritons are injected to the system with two different horizontal velocities corresponding to the incidence angles of the two laser beams. We adjust the intensities and incidence angles of two laser beams as well as the positions of laser spots on the surface of a microcavity sample in order to achieve the regime of shock-wave formation. The two beams originate from the same laser, so that their frequencies coincide, and the required phase matching condition between the two polariton flows created in this scheme is imposed with use of an optical delay line. The high quality factor of a microcavity ensures that polariton losses are negligible on the timescale of shock-wave formation and analogue Hawking emission, placing the system in the quasi‑conservative regime. The dynamics of the condensate wavefunction $\psi(x,t)$ are therefore well described by the quasi‑conservative Gross–Pitaevskii equation
\begin{align}\label{eq1}
i\hbar \frac{\partial \psi}{\partial t}=\left[-\frac{\hbar^2}{2m^{*}} \frac{\partial^{2}}{\partial x^{2}}+g|\psi|^2\right]\psi.
\end{align}
By redefining time, space, and polariton field in characteristic units,
\[
\tau = \frac{t}{\tau_{0}}, \qquad 
\xi = \frac{x}{R_{x}}, \qquad 
u = \frac{\psi}{\psi_{0}},
\]
the governing equation takes the following dimensionless form:
\begin{align}\label{eq2}
i \frac{\partial u}{\partial \tau} = -\frac{1}{2}\frac{\partial^{2} u}{\partial \xi^{2}}+\alpha|u|^2 u,
\end{align}
here $\alpha=g\psi_{0}^{2}\tau_{0}/\hbar$ and $\tau_{0}=m^{*}R_{x}^{2}/\hbar$. By applying the Madelung transformation $u = \sqrt{\rho}\, e^{i\phi},  \partial_\xi \phi = v,$ where $\rho$ denotes the density and $v$ represents the fluid velocity, Eq.~(\ref{eq2}) can be recast into the hydrodynamic form:
\begin{align}\label{eq3}
\alpha \rho_{\xi} + v v_{\xi} + v_{\tau}
= \frac{1}{4} \partial_\xi
\left(
\frac{\rho_{\xi\xi}}{\rho}
- \frac{\rho_{\xi}^2}{2\rho^2}
\right),
\end{align}
\begin{align}\label{eq4}
\rho_{\tau} + v \rho_{\xi} + \rho v_{\xi} = 0.
\end{align}
Under the traveling-wave transformation 
$\rho(\xi,\tau)=\rho(\zeta)$ and $v(\xi,\tau)=v(\zeta)$ 
with $\zeta=\xi - U \tau$, 
Eqs.~(\ref{eq3})-(\ref{eq4}) admit a periodic solution that can be written as
\begin{align}
\rho(\zeta) = \rho_2 - (\rho_2 - \rho_3)\, 
\mathrm{cn}^2\!\left[\sqrt{\alpha(\rho_1 - \rho_3)}\,(\zeta - \zeta_0),\, m\right],\nonumber 
\end{align}
\begin{align}\label{eq5}
v(\zeta) = U + \frac{C_1}{\rho(\zeta)},
\end{align}
the integration constant and the modulus are given by
\begin{align}\label{eq6}
C_1 &= \sqrt{\alpha\, \rho_1\rho_2\rho_3}, \qquad
m = \frac{\rho_2 - \rho_3}{\rho_1 - \rho_3},
\end{align}
where the parameters satisfy $\rho_1 > \rho_2 > \rho_3$.
In the dispersionless limit, Eqs.~(\ref{eq3})-(\ref{eq4}) can be diagonalized into the Whitham equations corresponding to the rarefaction waves by introducing the Riemann invariants
\(\lambda_{\pm} =-\frac{1}{2} v \pm \sqrt{\alpha \rho}\) 
and the characteristic velocities 
\(V_{\pm} = v \mp \sqrt{\alpha \rho}\),
\begin{align}\label{eq7}
\partial_{\tau}\lambda_{\pm} + V_{\pm}\partial_{\xi}\lambda_{\pm} = 0.
\end{align}
Taking into account the full dispersive effects, the system dynamics are described by the Whitham modulation equations expressed through four Riemann invariants $\lambda_1, \lambda_2, \lambda_3,$ and $\lambda_4$ \cite{el1995decay}. The relations that express the physical parameters of the periodic solution through the Riemann invariants are summarized in the Supplementary Information (SI).

In this work, we consider a Riemann-type initial value problem, where the density $\rho(\xi,0)> 0$ and velocity $v(\xi,0)$ are prescribed as piecewise constant functions:
\begin{align}\label{eq11}
\rho(\xi,0)=
\begin{cases}
\rho_0, & \xi<0,\\
\rho_R, & \xi>0,
\end{cases}
\qquad
v(\xi,0)=
\begin{cases}
v_0, & \xi<0,\\
v_R, & \xi>0.
\end{cases}
\end{align}
Correspondingly, the initial data of the Whitham equations can be expressed  as
\begin{align}\label{eq12}
\lambda_{+}(\tau=0)=
\begin{cases}
\lambda_+^L, & \xi<0,\\
\lambda_+^R, & \xi>0, 
\end{cases}
\quad
\lambda_{-}(\tau=0)=
\begin{cases}
\lambda_-^L,& \xi<0\\
\lambda_-^R, & \xi>0.
\end{cases}
\end{align}
Introducing the self-similar variable $\beta = \xi/\tau$, Eq.~(\ref{eq7}) admits two types of rarefaction waves, while Eq.~(S1) admits two types of dispersive shock waves \cite{gong2023whitham,el2016d}. These self-similar solutions correspond to the nontrivial simple waves in the Riemann problem. In general, the solution to the Riemann initial-value problem for Eq.~(\ref{eq2}) consists of three fundamental types of simple waves: plateaus, rarefaction waves, and dispersive shock waves \cite{gong2023whitham}.

\begin{figure}[h]
\centering
\includegraphics[width=0.9\linewidth]{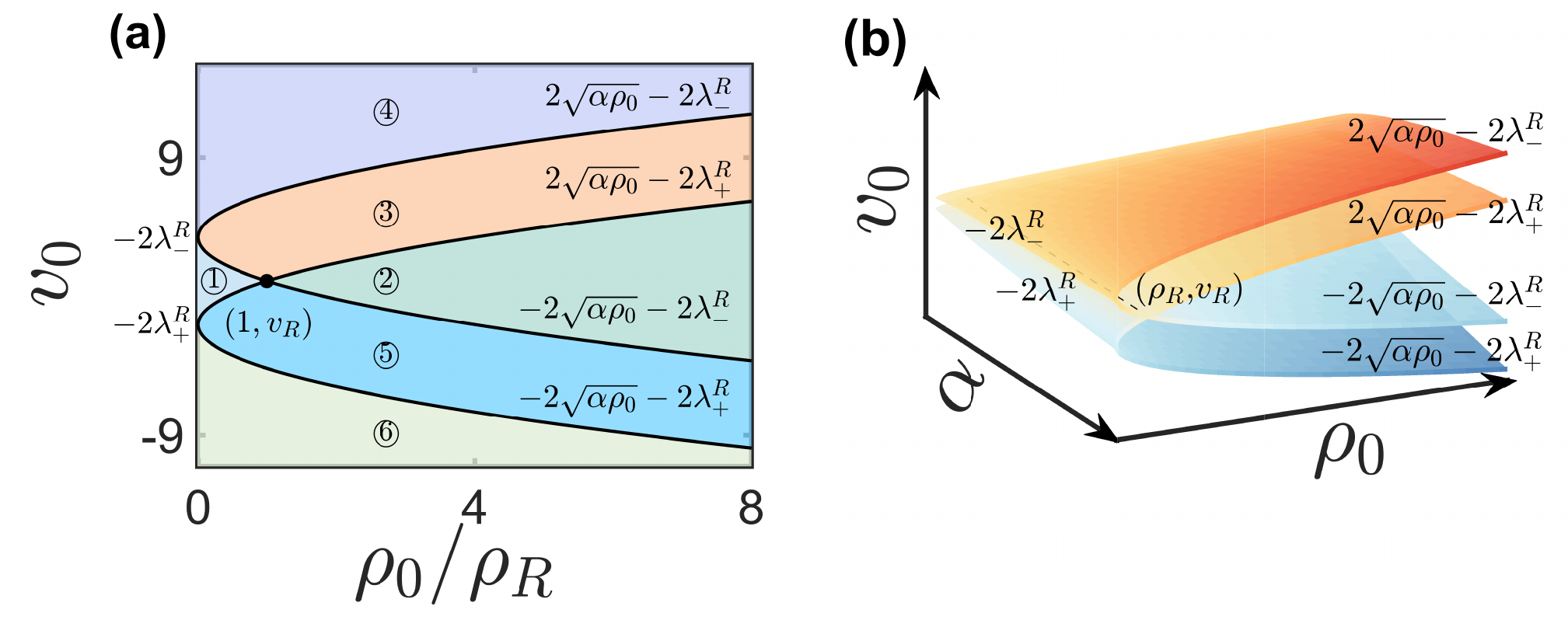}
\caption{\footnotesize  (a) Classification diagram of wave regions in the $(\rho_0, v_0)$ plane. The different wave regions are labeled with circled numbers.
(b) Schematic diagram showing the existence regions of different wave solutions in the $(\rho_0, \alpha, v_0)$ space, which vary with the parameter $\alpha$. The black dashed line represents the positions of all intersection points $(\rho_R, v_R)$ for different values of $\alpha$.
 } \label{fig1}
\end{figure}

By fixing the right initial state $(\rho_R, v_R)$, the corresponding values of the Riemann invariants on the right side are determined, which allows the identification of six distinct wave regions, as shown in Fig.~\ref{fig1}(a). The two parabolic curves intersect at $(\rho_R, v_R)$. In addition, the regions are labeled with circled numbers, and their classification is related to the ordering of the four initial values of the Riemann invariants \cite{jenkins2015r}. When considering different strengths of the repulsive interaction $\alpha$ between polaritons, the corresponding wave region diagrams are distinct. Fig.~\ref{fig1}(b) shows that as the repulsive interaction increases (in the direction of the black arrow), the regions of existence for the different waves expand  in the $(\rho_0, v_0)$ space. Additionally, for each value of $\alpha$, the intersection points of the corresponding region diagrams always coincide at $(\rho_R, v_R)$, illustrating the invariance of the right boundary conditions.

\begin{figure}[h]
\centering
\includegraphics[width=0.9\linewidth]{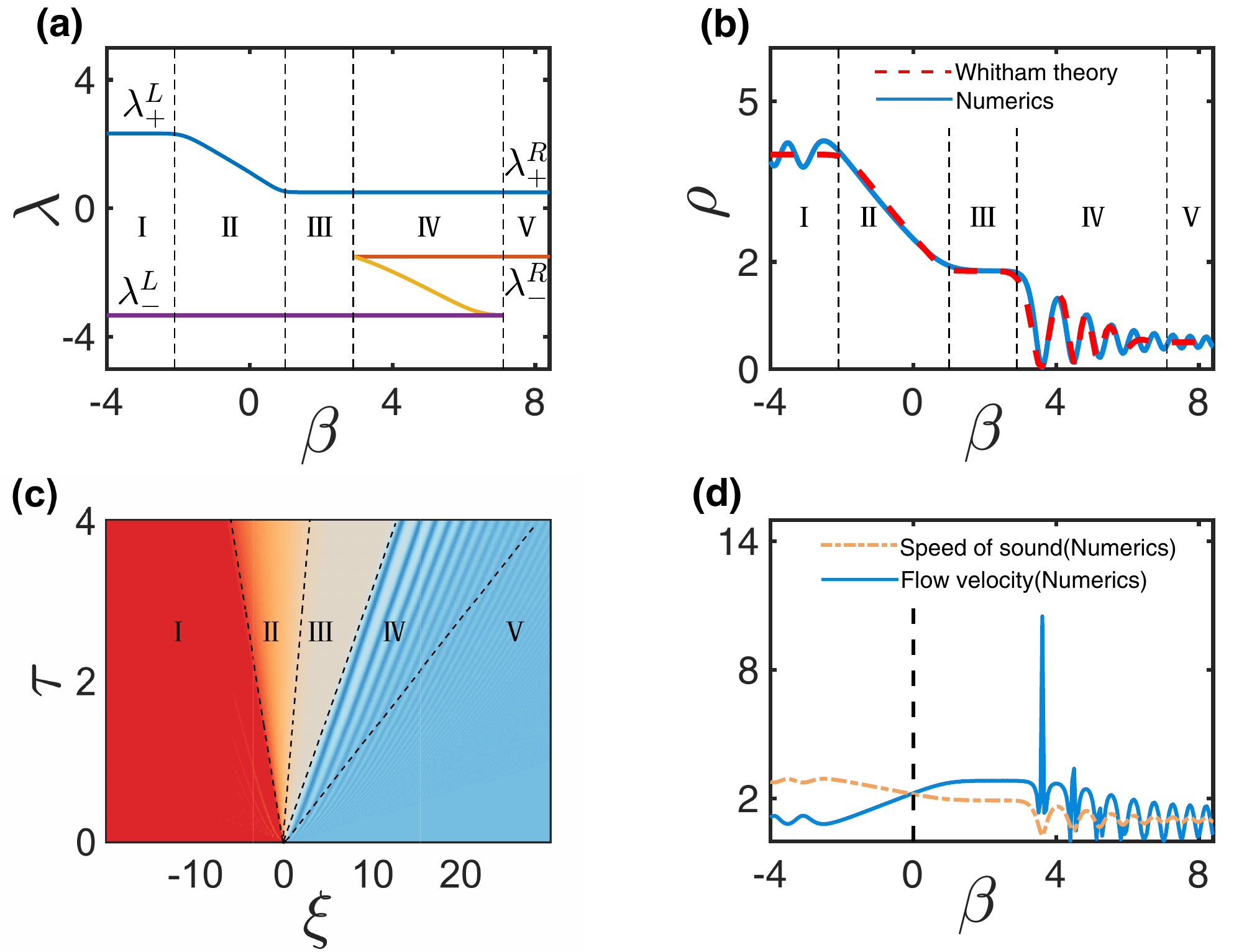}
\caption{\footnotesize (a) Distribution of Riemann invariants; (b) Waveform structure of the density $\rho$; (c) Spatiotemporal evolution of the density $\rho$ up to $\tau=4$; In (b), the red dashed and blue solid lines represent the theoretical and numerical results, respectively. (d) Comparison between numerical results of flow velocity and sound speed;  In (d), the orange dash-dotted line represents the sound speed, and the blue solid line represents the flow velocity. (a), (b), and (d) present the results at $\tau = 2$. The parameters are $\alpha=2$, $(\rho_R, v_R)=(0.5,1)$, and $(\rho_0, v_0)=(4,1)$.
} \label{fig3}
\end{figure}

We focus on Region (2), where the corresponding Riemann invariants satisfy the constraint $ \lambda_+^L > \lambda_+^R> \lambda_-^R >\lambda_-^L $. Following previous studies have identified two basic wave types: rarefaction waves and dispersive shock waves. the corresponding distributions of the Riemann invariants for the initial conditions $\alpha = 2$, $(\rho_R, v_R) = (0.5, 1)$, and $(\rho_0, v_0) = (4, 1)$ are shown in Fig.~\ref{fig3}(a). Fig.~\ref{fig3}(b) displays the distribution of the density $\rho$, obtained from the Whitham theory (red dashed line) and numerical simulation (blue solid line). The corresponding flow velocity $v$, obtained in the same way, is shown in Fig.~S1(a). The five regions labeled $\mathrm{I}$-$\mathrm{V}$ in Figs.~\ref{fig3}(a)-\ref{fig3}(c) correspond, from left to right, to plateau, rarefaction wave, plateau, shock wave and plateau structures. The numerical results agree well with predictions of the Whitham modulation theory. From the evolution of the density $\rho$ shown in Fig.~\ref{fig3}(c), regions $\mathrm{II}$, $\mathrm{III}$, and $\mathrm{IV}$ are observed to expand gradually over time. In particular, region $\mathrm{IV}$ exhibits pronounced oscillations that are characteristic of a dispersive shock wave.
For the elliptic periodic-wave solution presented in Eq.~\ref{eq5}, 
as the modulus $m \to 1$ ($\rho_1 \to \rho_2$), 
the solution gradually transforms into a localized solitary-wave form,
\begin{align}\label{eq10}
\rho(\zeta)
= \rho_3 + (\rho_2 - \rho_3)\,
\tanh^2\!\left(\sqrt{\alpha (\rho_2 - \rho_3)}\,(\zeta - \zeta_0)\right).
\end{align}
This limit represents the smooth transition of the periodic structure into a dark soliton, which coincides with the waveform located at the left boundary of region~$\mathrm{IV}$ in Fig.~\ref{fig3}(b). When the modulus $m \to 0$ ($\rho_3 \to \rho_2$), the oscillation amplitude tends to vanish and the solution reduces to a uniform state $\rho = \rho_2$, corresponding to the uniform background predicted by the Whitham theory at the right edge of region~$\mathrm{IV}$ in Fig.~\ref{fig3}(b). 
Fig.~\ref{fig3}(d)  provides a comparison between the flow velocity and the sound speed.
The flow velocity exceeds the sound speed in the shock-wave region, clearly signaling the breakdown of superfluidity. It is also evident that locally supersonic flow appears in the rarefaction wave region.

\paragraph*{Acoustic horizon and analogue black hole ---}
To identify and characterize the analogue black hole behavior, 
we analyze small perturbations on top of the slowly varying 
background solution of Eq.~\ref{eq1}. 
Writing the field as
\begin{equation}
\psi(x,t)=\left[\sqrt{\rho_0(x,t)}+\delta\rho(x,t)\right]
e^{i[\theta_0(x,t)+\delta\theta(x,t)]},
\end{equation}
and linearizing in $\delta\rho$ and $\delta\theta$, 
we obtain the coupled equations
\begin{subequations}
\begin{align}
&\partial_t \delta\rho 
+ \partial_x(\rho_0\,\delta v + v_0\,\delta\rho) = 0, \\
&\partial_t \delta v 
+ v_0 \partial_x \delta v
+ \frac{g}{m^*}\partial_x\!\left(\frac{\delta\rho}{\rho_0}\right)
- \frac{\hbar^2}{4m^{*2}}\partial_x^3
\!\left(\frac{\delta\rho}{\rho_0}\right)=0,
\end{align}
\end{subequations}
where $v_0=(\hbar/m^*)\partial_x \theta_0$ and 
$c_0(x,t)=\sqrt{g\rho_0/m^*}$ denote the local flow and sound speeds, respectively.  
Neglecting the quantum pressure term in the long-wavelength limit 
($k\xi\ll1$ with $\xi=\hbar/\sqrt{2m^*g\rho_0}$ the healing length), 
one can eliminate $\delta v$ to obtain a single wave equation for the phase fluctuation $\phi=\delta\theta$:
\begin{widetext}
\begin{equation}
\partial_t\!\left[\frac{\rho_0}{c_0^2}(\partial_t\phi+v_0\partial_x\phi)\right]
-\partial_x\!\left[\rho_0\,(c_0^2-v_0^2)\partial_x\phi 
- \rho_0\,v_0(\partial_t\phi+v_0\partial_x\phi)\right]=0.
\end{equation}
\end{widetext}
This equation can be recast in covariant form 
as a massless Klein-Gordon equation
\begin{equation}
\frac{1}{\sqrt{-g}}\,
\partial_\mu\!\left(\sqrt{-g}\,g^{\mu\nu}\partial_\nu \phi\right)=0,
\end{equation}
where $g^{\mu\nu}$ is the acoustic metric:
\begin{equation}
g_{\mu\nu}=\frac{\rho_0}{c_0}
\begin{pmatrix}
-(c_0^2-v_0^2) & -v_0\\[4pt]
-v_0 & 1
\end{pmatrix},
\qquad 
\mu,\nu\in\{t,x\}.
\end{equation}
Therefore, long-wavelength Bogoliubov excitations propagate as scalar fields in a curved $(1+1)$-dimensional spacetime with line element
\begin{equation}
ds^2 = g_{\mu\nu}dx^\mu dx^\nu 
= \frac{\rho_0}{c_0}\Big[-(c_0^2-v_0^2)dt^2 
- 2v_0\,dt\,dx + dx^2 \Big].
\end{equation}

An acoustic horizon occurs where the local Mach number $M=|v_0|/c_0$ crosses unity. At this point, the metric coefficient $g_{tt}=-(c_0^2-v_0^2)$ vanishes, producing a unidirectional causal structure analogous to that of a gravitational black hole: subsonic excitations ($|v_0|<c_0$) can propagate both upstream and downstream, whereas supersonic excitations ($|v_0|>c_0$) are convected downstream and cannot escape across the horizon. From the above, the following conclusion can be drawn: analogue Hawking radiation originates in the vicinity of the event horizon, where the flow velocity equals the local speed of sound. Subsequently, the resulting fluctuations propagate almost freely through the medium. Therefore, for the radiation process, the effective analogue gravitational field at the horizon is of primary importance. It is evident from general relativity, or directly from the field equations, that the gravitational field is proportional to the derivatives of the metric components. Consequently, one can introduce the concept of surface gravity
\begin{align}
\kappa=\frac{1}{2c_h}\frac{\partial}{\partial x}(v_0^2-c_0^2)\bigg|_{x=x_h},
\end{align} where $c_h=c_0(x_h,t)$. The production rate of fluctuations is proportional to the magnitude of this quantity. Moreover, most intriguing is the fact that the fluctuation spectrum follows a thermal distribution corresponding to the Hawking temperature, which is given by the formula:
\begin{align}
    T_H=\frac{\hbar \kappa}{2\pi k_B}.
    \label{eqHT}
\end{align}
This is a consequence of the fact that the spacetime geometry near the horizon satisfies the Kubo-Martin-Schwinger condition. In essence, this condition implies that the effective geometry becomes periodic under a Wick rotation into imaginary time (see SI). This periodicity, in turn, forces the spectrum to become discrete. For a physical system, this periodicity dictates the spacing of the Matsubara frequencies, which is intrinsically linked to the temperature.
This equation provides a meaningful effective temperature characterizing the strength of horizon-induced pair creation. Benefiting from the high velocities of both sound and flow in the exciton-polariton system ($10^6 \ \rm m/s$), the Hawking temperature calculated using Eq.~\ref{eqHT} may reach $1.14 \rm \ K$, which is $10^{10}$ higher than previously reported $0.351 \rm\ nK$ in the cold atom system\cite{munoz2019observation}, where the velocity is around $10^{-4} \ \rm m/s$. The corresponding characteristic wavelength of Hawking radiation is $\lambda=17.5 \rm\ \mu m$, indicating the possibility for experimental observation of Hawking temperature in a micrometer-sized sample. The occupation number of emitted modes follows the approximate Bose-Einstein distribution
\begin{equation}
n_\omega = \frac{1}{e^{\hbar\omega/(k_B T_H)}-1},
\label{eq:bed}
\end{equation}
which defines the analogue Hawking spectrum.
$T_H$ represents a geometric scale of the analogue of surface gravity that governs the exponential redshift of Bogoliubov modes near the horizon. If quantum fluctuations are included, the spectral distribution of the emitted quasiparticles follows an approximately thermal law with an effective temperature set by $T_H$. In this sense, $T_H$ characterizes the strength and frequency scale of horizon-induced mode conversion, not the physical temperature of the condensate.

\begin{figure}
    \centering
    \includegraphics[width=0.9\linewidth]{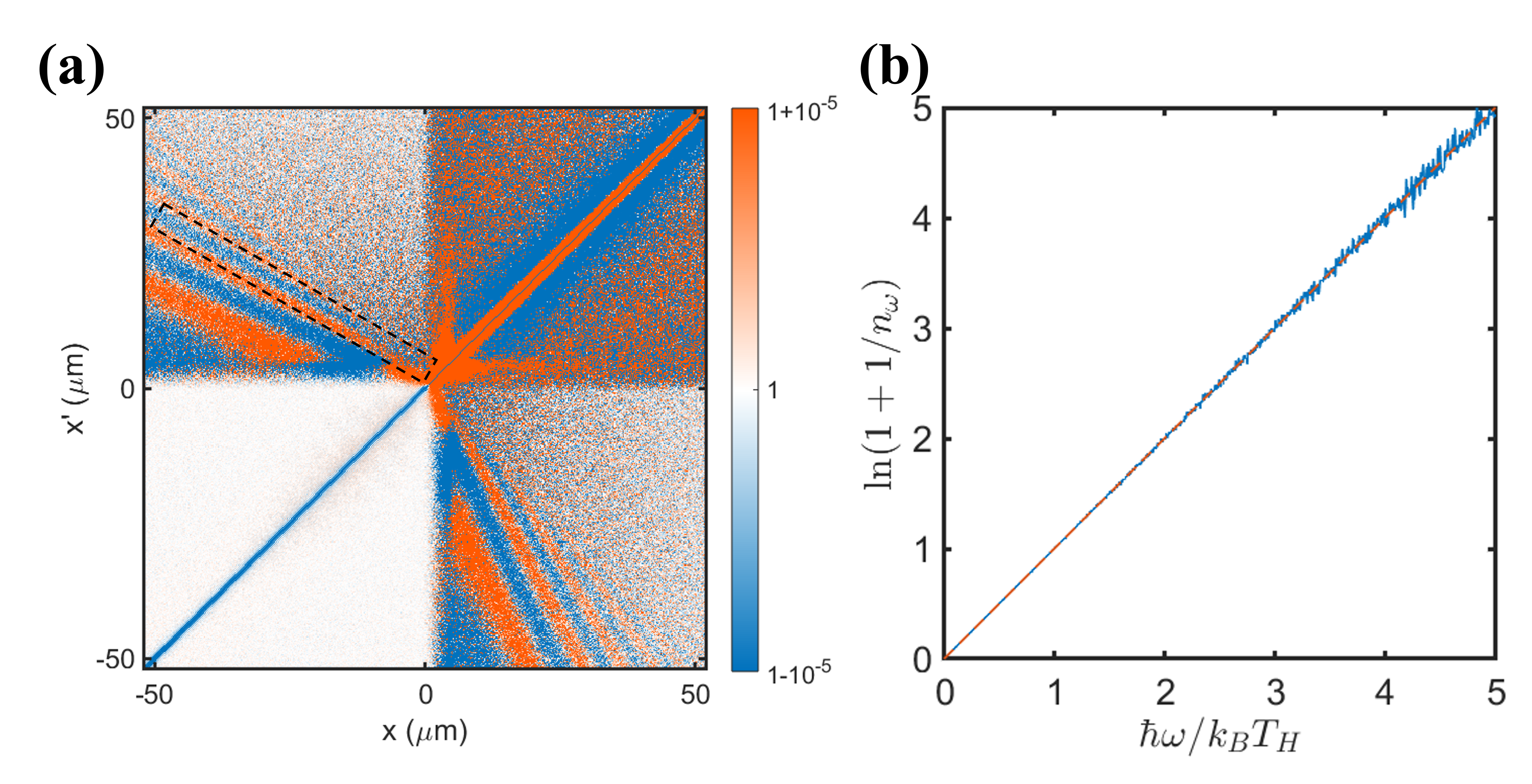}
    \caption{(a) Second order correlation function $g^{(2)}(x,x^\prime)$. The off-diagonal curve shows a negative correlation as a signature of the quantum noise passing through the acoustic horizon. (b) The distributions of the Hawking radiation calculated from the correlation function in the dashed box of (a) (blue solid curve), and from the surface gravity (red dashed curve). Parameters: $m^*=5\times10^{-5}m_e$, $g\rho_0=0.1\ \rm meV$, $\phi(x<0,t=0)=2\sqrt{\rho_0}e^{-ik_Lx}$, $\phi(x\geq0,t=0)=\sqrt{\rho_0}e^{-ik_Rx}$, $k_L=0.51\ \rm \mu m^{-1}$, $k_R=0.78\ \rm \mu m^{-1}$}
    \label{fig:g2}
\end{figure}

Linear perturbations in the stationary background take the Bogoliubov form 
$\delta\psi \!\propto\! u(x)e^{-i\omega t}+v^*(x)e^{i\omega t}$,
which satisfy the coupled eigenvalue equations
\begin{subequations}
\begin{align}
\hbar\omega\,u &= 
\left[-\frac{\hbar^2}{2m^*}\partial_x^2 - \mu 
+ 2g\rho_0\right]u + g\rho_0\,v, \\
-\hbar\omega\,v &=
\left[-\frac{\hbar^2}{2m^*}\partial_x^2 - \mu 
+ 2g\rho_0\right]v + g\rho_0\,u,
\end{align}
\end{subequations}
with chemical potential $\mu=g\rho_0+ m^*v_0^2/2$. The dispersion relation reads
\begin{equation}
(\omega - v_0 k)^2 = c_0^2 k^2 
+ \frac{\hbar^2 k^4}{4m^{*2}},
\end{equation}
whose linear part reproduces the acoustic modes and whose quartic term accounts for the dispersion. At the horizon, counter-propagating (negative-norm) modes can be converted into co-propagating (positive-norm) modes, producing correlated quasiparticle pairs with opposite frequencies in the comoving frame. To reveal these effects in simulations or experiments, we analyze the density–density correlation function. The equal-time correlation function,
\begin{equation}
g^{(2)}(x,x';t)
= \frac{\langle |\psi(x,t)|^2|\psi(x',t)|^2\rangle}
{\langle |\psi(x,t)|^2\rangle
 \langle |\psi(x',t)|^2\rangle},
\end{equation}
is obtained by ensemble averaging over stochastic realizations within the truncated Wigner method. As shown in Fig.~\ref{fig:g2} (a), the distinct off-diagonal correlation stripes aligned along the characteristic trajectories of the emitted quasiparticles indicate the presence of Hawking pair production. The Hawking temperature can also be extracted from the correlation function\cite{munoz2019observation}. Define the density-density correlation of the Hawking pair $G^{(2)}(x,x')=\langle \delta n(x)\,\delta n(x') \rangle=\langle n(x)\rangle\langle n(x^{\prime})\rangle
[g^{(2)}(x,x^{\prime})-1]$. After the double Fourier transform $\widetilde G^{(2)}(k,k')=\int dx\,dx'\,G^{(2)}(x,x')\,e^{-i(kx+k'x')}$ in the region of the dashed box in Fig.~\ref{fig:g2} (a), one will find the
\begin{equation}
\big|\widetilde G^{(2)}(k,k')\big|^2\propto(n_\omega+1)n_\omega.
\label{eq:g2kn}
\end{equation}
In Fig.~\ref{fig:g2} (b), we show $\ln(1+1/n_\omega)$ calculated from Eq.~\ref{eq:g2kn} along the off-diagonal correlation curve (blue solid curve) and from Eq.~\ref{eq:bed} (red dashed line) within the energy range from zero to $5k_BT_H$ with $T_H=1.14 \rm\ K$. It is observed that the distribution calculated from the Fourier transform of the Hawking pair density-density correlation function $G^{(2)}(x,x')$ is in agreement with the Bose-Einstein distribution given by Eq.~\ref{eq:bed}, which quantitatively confirms the Hawking temperature and Hawking radiation.

\paragraph*{Conclusion ---}
We have demonstrated that dispersive shock waves in a quasi-conservative exciton-polariton condensate generated by a double-beam resonant pumping can spontaneously form stable acoustic black holes without the need for externally engineered potentials or pumping profiles. Our work establishes shock waves as a robust and intrinsic mechanism for creating acoustic horizons in quantum fluids of light. It bridges the study of nonlinear dispersive hydrodynamics with analogue gravity, offering a new pathway to explore quantum effects in curved effective spacetimes within highly controllable polariton systems. Future experiments measuring spatiotemporal spectra and Hanbury Brown–Twiss correlations in optically generated polariton shocks could directly observe the predicted analogue Hawking radiation and horizon dynamics.

AVK acknowledges support from Saint Petersburg State University (Research Grant No. 125022803069-4) and from the Innovation Program for Quantum Science and Technology (No. 2021ZD0302704). HJL acknowledges National Natural Science Foundation of China (12375006) and postdoctoral fellowship of Zhejiang Normal University (YS304123952).

JC and JW contributed equally to this work.

\bibliography{apssamp}

\end{document}


\title{Supplementary Information}

\author{Junhui Cao$^{1}$}%
\email{tsao.c@mipt.ru}
\author{Jinling Wang$^{2}$}
\author{Kirill Bazarov$^{1}$}%
\author{Chenqi Jin$^{2}$}
\author{Huijun Li$^{2,3}$}
\email{hjli@zjnu.cn}
\author{Anton Nalitov$^{1}$}%
\author{Alexey Kavokin$^{1,4}$}%
\email{a.kavokin@westlake.edu.cn}
\affiliation{$^{1}$Abrikosov Center for Theoretical Physics, Moscow Center for Advanced Studies, Moscow 141701, Russia.
\\$^{2}$Institute of Nonlinear Physics and Department of Physics, Zhejiang Normal University, Jinhua, 321004 Zhejiang, China\\
$^{3}$Zhejiang Institute of Photoelectronics, Jinhua, Zhejiang 321004, China\\
$^{4}$Department of Physics, St. Petersburg State University, University Embankment, 7/9, St. Petersburg, 199034, Russia}




\begin{abstract}
           
\end{abstract}

\maketitle

\section{WHITHAM MODULATION EQUATIONS AND RELATIONS BETWEEN THE RIEMANN INVARIANTS AND THE PHYSICAL PARAMETERS}\label{I}
Starting from Eqs.~(3)-(4), the corresponding Whitham modulation equations can be expressed as
\begin{align}\label{eq8}
&\frac{\partial \lambda_i}{\partial \tau } + v_i(\lambda_1, \lambda_2, \lambda_3, \lambda_4) \frac{\partial \lambda_i}{\partial \xi} = 0, \quad \lambda_1 > \lambda_2 > \lambda_3 > \lambda_4,\nonumber\\ 
& i = 1, 2, 3, 4,
\end{align}
where the characteristic velocities \( v_i(\lambda_i) \) take the form:
\begin{align}\label{eq9}
v_1 &= -\frac{1}{2} \sum_{i=1}^{4} \lambda_i - \frac{(\lambda_1 - \lambda_4)(\lambda_1 - \lambda_2) K(m)}{(\lambda_1 - \lambda_4) K(m) + (\lambda_4 - \lambda_2) E(m)}, \nonumber\\
v_2 &= -\frac{1}{2} \sum_{i=1}^{4} \lambda_i + \frac{(\lambda_2 - \lambda_3)(\lambda_1 - \lambda_2) K(m)}{(\lambda_2 - \lambda_3) K(m) + (\lambda_3 - \lambda_1) E(m)}, \nonumber\\
v_3 &= -\frac{1}{2} \sum_{i=1}^{4} \lambda_i - \frac{(\lambda_2 - \lambda_3)(\lambda_3 - \lambda_4) K(m)}{(\lambda_2 - \lambda_3) K(m) + (\lambda_4 - \lambda_2) E(m)}, \nonumber\\
v_4 &= -\frac{1}{2} \sum_{i=1}^{4} \lambda_i + \frac{(\lambda_1 - \lambda_4)(\lambda_3 - \lambda_4) K(m)}{(\lambda_1 - \lambda_4) K(m) + (\lambda_3 - \lambda_1) E(m)},
\end{align}
here \( m = [(\lambda_1 - \lambda_2)(\lambda_3 - \lambda_4)]/[(\lambda_1 - \lambda_3)(\lambda_2 - \lambda_4)]\); the Riemann invariants \( \lambda_1, \lambda_2, \lambda_3, \lambda_4 \) vary slowly with \(\xi\) and \(\tau\), with \(K(m)\) and \(E(m)\) denoting the complete elliptic integrals of the first and second kinds, respectively. In addition, the parameters \(\rho_1, \rho_2, \rho_3\) can be expressed in terms of the Riemann invariants as
\begin{align}\label{eq10}
\rho_1 &= \tfrac{1}{4 \alpha } (\lambda_1 + \lambda_2 - \lambda_3 - \lambda_4)^2, \nonumber\\
\rho_2 &= \tfrac{1}{4 \alpha} (\lambda_1 - \lambda_2 + \lambda_3 - \lambda_4)^2, \nonumber\\
\rho_3 &= \tfrac{1}{4 \alpha} (\lambda_1 - \lambda_2 - \lambda_3 + \lambda_4)^2,\nonumber\\
U &=-\tfrac{1}{2}(\lambda_1 + \lambda_2+ \lambda_3+ \lambda_4).
\end{align}

\section{DYNAMICAL BEHAVIORS IN THE REMAINING REGIONS OF Fig. 2}\label{II}

\subsection{Region (1)}
Similar to Fig.~3, only one dispersive shock wave is generated in this region, and the associated Riemann invariants obey the ordering $\lambda_+^R > \lambda_+^L > \lambda_-^L > \lambda_-^R$. The five regions labeled $\mathrm{I}$-$\mathrm{V}$ in Figs.\ref{fig2}(a)-\ref{fig2}(d) correspond, from left to right, to plateau, shock wave, plateau, rarefaction wave, and plateau structures, with pronounced oscillations observed in region~II.
Figs~\ref{fig2}(e) and \ref{fig2}(f) present a comparison between the flow velocity and the sound speed. It is found that the flow velocity exceeds the sound speed in region~$\mathrm{II}$, indicating a locally supersonic flow.

\subsection{Region (3)}
In this region, only shock waves and plateau wave structures are present. Fig.~\ref{fig4}(a) shows that the corresponding Riemann invariants for this wave satisfy the constraint $\lambda_+^R >\lambda_+^L > \lambda_-^R>  \lambda_-^L $. The five regions labeled $\mathrm{I}$-$\mathrm{V}$ in Figs.~\ref{fig4}(a)-\ref{fig4}(d) correspond, from left to right, to plateau, shock wave, plateau, shock wave, and plateau structures. In this case, the plateau in region~$\mathrm{III}$ connects the two shock waves located in regions~$\mathrm{II}$ and ~$\mathrm{IV}$. Figs.~\ref{fig4}(e)-\ref{fig4}(f) demonstrate that the flow in region~$\mathrm{II}$ is generally supersonic, exhibiting a transition to subsonic flow only near the soliton edge in an averaged sense, whereas in region~$\mathrm{IV}$ it becomes locally supersonic close to the soliton edge.

\subsection{Region (4)}
In this region, two shock waves are present, and the associated Riemann invariants obey the ordering $\lambda_+^R> \lambda_-^R >\lambda_+^L >  \lambda_-^L $. Moreover, due to the sufficiently large velocity $v_0$, a periodic wave structure appears in region $\mathrm{III}$ of Figs.~\ref{fig5}(a)-\ref{fig5}(c). The wave structure originates from the overlap of two dispersive shock waves, leading to a two-phase modulation solution (non–self-similar). Under self-similar conditions, the two-phase solution degenerates into a one-phase solution.

\subsection{Region (5)}
In this region, no shock waves are present. Fig.~\ref{fig6}(a) shows that the corresponding Riemann invariants for this wave satisfy the constraint $\lambda_+^L >\lambda_+^R>\lambda_-^L > \lambda_-^R $. The five regions labeled $\mathrm{I}$-$\mathrm{V}$ in Figs.~\ref{fig6}(a)-\ref{fig6}(c) correspond, from left to right, to plateau, rarefaction wave, plateau, rarefaction wave, and plateau structures.

\subsection{Region (6)}
In this region, no shock waves are present. Fig.~\ref{fig7}(a) shows that the corresponding Riemann invariants for this wave satisfy the constraint $\lambda_+^L>\lambda_-^L >\lambda_+^R > \lambda_-^R $. The five regions labeled $\mathrm{I}$-$\mathrm{V}$ in Figs.~\ref{fig7}(a)-\ref{fig7}(c) correspond, from left to right, to plateau, rarefaction wave, plateau, rarefaction wave, and plateau structures. In contrast to Fig.\ref{fig6}, region $\mathrm{III}$ in Fig.~\ref{fig7} exhibits a vacuum point where the density $\rho$ is zero. For solutions composed of rarefaction waves and plateaus, two distinct flow regimes may arise: a fully subsonic regime, as illustrated in Fig.~\ref{fig6}, and a mixed regime featuring coexisting subsonic and supersonic regions, as shown in Fig.~\ref{fig7}.

\section{INFLUENCE OF SELF-INTERACTION STRENGTH}\label{III}
To illustrate the effect of self-interaction strength, three representative cases are considered, which correspond to $\alpha =0.5$, $2$, and $4$ and are shown in Figs.~\ref{fig8}, \ref{fig2}, and \ref{fig9}, respectively, while all other parameters are held constant. By comparing the results under different self-interaction strengths, we find that increasing self-interaction strength leads to a larger wavenumber in the shock-wave region. Moreover, both the waveform structure and the density evolution indicate that regions $\mathrm{II}$,$\mathrm{III}$, and $\mathrm{IV}$ progressively expand as the self-interaction becomes stronger. Although the self-interaction strength varies, the comparison between the flow velocity and the sound speed shows a consistent transition from a supersonic to a subsonic regime near the soliton edge.

\section{Truncated Wigner Approximation}

In order to characterize quantum fluctuations and two-point density correlations beyond the mean-field theory, we employ the truncated Wigner approximation (TWA) method. This semiclassical phase-space method allows us to include the leading quantum noise on top of the Gross-Pitaevskii dynamics and is well suited for weakly interacting Bose fields with large occupation numbers.

We consider a bosonic field operator $\hat{\psi}(x,t)$ satisfying the usual equal-time commutation relation
\begin{equation}
[\hat{\psi}(x),\hat{\psi}^\dagger(x')] = \delta(x-x').
\end{equation}
Within the Wigner representation, quantum expectation values of symmetrically ordered operator products are mapped onto ensemble averages over a complex stochastic field $\psi_W(x,t)$:
\begin{equation}
\langle \{ \hat{\psi}^\dagger(x,t)\hat{\psi}(x',t) \} \rangle
= \langle \psi_W^*(x,t)\psi_W(x',t) \rangle,
\end{equation}
where $ \langle \psi_W^*(x,t)\psi_W(x',t) \rangle$ denotes averaging over the Wigner representations via stochastic realizations.

The time evolution of $\psi_W$ is governed by a stochastic Gross-Pitaevskii equation. In the truncated Wigner approximation, third- and higher-order derivatives in the corresponding Fokker-Planck equation are neglected, resulting in a classical field evolution with quantum noise encoded in the initial condition. For an initially coherent or weakly depleted condensate, the Wigner field at $t=0$ is sampled as
\begin{equation}
\psi_W(x,0) = \psi_0(x) + \frac{1}{\sqrt{2}}\left[\xi_1(x) + i\xi_2(x)\right],
\end{equation}
where $\psi_0(x)$ is the mean-field solution and $\xi_{1,2}(x)$ are independent real Gaussian random variables satisfying
\begin{equation}
\langle \xi_\alpha(x)\xi_\beta(x') \rangle = \delta_{\alpha\beta}\delta(x-x').
\end{equation}
This prescription corresponds to half a quantum of vacuum noise per mode, as required by the Wigner representation.

In numerical simulations, space is discretized on a uniform grid with spacing $\Delta x$, and the field operators are replaced by lattice operators
\begin{equation}
\hat{\psi}(x_j) \rightarrow \frac{\hat{a}_j}{\sqrt{\Delta x}},
\end{equation}
such that $[\hat{a}_j,\hat{a}_k^\dagger] = \delta_{jk}$. Accordingly, the Wigner field on the lattice is sampled as
\begin{equation}
\psi_j(0) = \psi_{0,j} + \frac{1}{\sqrt{2\Delta x}}(\eta_{1,j} + i\eta_{2,j}),
\end{equation}
where $\eta_{1,j}$ and $\eta_{2,j}$ are independent real Gaussian random numbers with
\begin{equation}
\langle \eta_{\alpha,j}\eta_{\beta,k} \rangle = \delta_{\alpha\beta}\delta_{jk}.
\end{equation}
The subsequent time evolution of $\psi_j(t)$ is governed by the discretized Gross-Pitaevskii equation, and observables are obtained by averaging over many stochastic trajectories. The physical density operator on the lattice is
\begin{equation}
\hat{n}_j = \hat{\psi}^\dagger(x_j)\hat{\psi}(x_j)
= \frac{\hat{a}_j^\dagger \hat{a}_j}{\Delta x}.
\end{equation}
Within the Wigner representation, the corresponding expectation value is obtained as
\begin{equation}
\langle \hat{n}_j \rangle
= \frac{1}{\Delta x}
\left(
\langle |\psi_j|^2 \rangle - \frac{1}{2}
\right),
\end{equation}
where the $1/2$ term represents the standard Wigner correction arising from symmetric ordering.

The equal-time, normally ordered density-density correlation function is defined as
\begin{equation}
g^{(2)}(x_j,x_k)
=
\frac{
\langle \hat{n}_j \hat{n}_k \rangle
}{
\langle \hat{n}_j \rangle
\langle \hat{n}_k \rangle
}.
\end{equation}
For $j\neq k$, the normally ordered product reads
\begin{equation}
\langle \hat{n}_j \hat{n}_k \rangle
=
\frac{1}{\Delta x^2}
\left(
\langle |\psi_j|^2 |\psi_k|^2 \rangle
- \frac{1}{2}\langle |\psi_j|^2 \rangle
- \frac{1}{2}\langle |\psi_k|^2 \rangle
\right).
\end{equation}
Substituting the expressions for $\langle \hat{n}_j \rangle$ and $\langle \hat{n}_k \rangle$, the second-order correlation function in the truncated Wigner approximation takes the explicit discrete form
\begin{equation}
g^{(2)}(x_j,x_k)
=
\frac{
\langle |\psi_j|^2 |\psi_k|^2 \rangle
- \frac{1}{2}\langle |\psi_j|^2 \rangle
- \frac{1}{2}\langle |\psi_k|^2 \rangle
}{
\left(\langle |\psi_j|^2 \rangle - \frac{1}{2}\right)
\left(\langle |\psi_k|^2 \rangle - \frac{1}{2}\right)
}.
\label{eq:g2_TWA_lattice}
\end{equation}
Equation~(\ref{eq:g2_TWA_lattice}) is the expression used to compute the correlation map shown in Fig.~4 of the main text. For sufficiently high local densities, the Wigner corrections become negligible and $g^{(2)}\rightarrow 1$, as expected for a coherent superfluid background. Deviations of $g^{(2)}(x,x')$ from unity encode correlations induced by quantum fluctuations. In particular, the appearance of off-diagonal negative correlation stripes indicates the production of correlated quasiparticle pairs propagating along distinct characteristic trajectories. In the presence of an acoustic horizon, these correlations are naturally interpreted as the analogue of Hawking pair creation, with one partner propagating into the supersonic region and the other escaping into the subsonic flow. Fig.~\ref{fig:si_g2} shows the velocity fields and second-order correlation functions for systems in the purely subsonic (a,b) and purely supersonic (c,d) regimes. In both cases, no acoustic horizon is present, though for different reasons. In the subsonic regime, quasi-particle excitations can propagate in both directions relative to the lab frame, so no waves become trapped and excitations can escape. Consequently, the strong mode-stretching and causal separation necessary for Hawking radiation do not occur. In contrast, in the supersonic regime, the flow everywhere exceeds the local speed of sound, placing the entire region causally behind where a horizon would be, with no external free region into which Hawking partners could escape. An acoustic horizon and the associated simulated Hawking effect can only arise at a transonic interface, which mimics a spacetime event horizon by separating subsonic (exterior) from supersonic (interior) domains.

\section{Derivation of Hawking temperature}
We put into the metric $c_0^2-v_0^2\approx 2\kappa c_0(x_h-x)$, then
\begin{align}
ds^2 = \frac{\rho_0}{c_0}\Big[-2\kappa c_0(x_h-x)dt^2 - 2c_0dtdx + dx^2\Big].
\end{align}
To diagonalize the metric we perform the coordinate transformation, introducing a new time $t = t' -\frac{1}{2\kappa}\ln|x_h-x|$, so that $dt = dt' -\frac{1}{2\kappa(x_h-x)}dx$. Substituting this into the line element, the cross term $dt'dx$ vanishes, and in the new coordinates the metric becomes diagonal:
\begin{align}
ds^2 = -2\rho_0\kappa(x_h-x)dt'^2 + \frac{\rho_0}{c_0}\Big[1+\frac{c_0}{2\kappa (x_h-x)}\Big]dx^2 .
\end{align}
Near the horizon $x\to x_h$, where $(x_h-x)\ll 1$, the term $1/(x_h-x)$ dominates and
\begin{align}
ds^2=-2\kappa\rho_0(x_h-x)dt'^2 
+ \frac{\rho_0 }{2\kappa (x_h-x)}dx^2 .
\end{align}
Defining a radial coordinate $r$ by
\begin{align}
r = 2\sqrt{\frac{\rho_0}{2\kappa}}\sqrt{x_h-x},
\quad \text{and} \quad
dr = \sqrt{\frac{\rho_0}{2\kappa}}\frac{dx}{\sqrt{x_h-x}},
\end{align}
the metric simplifies to
\begin{align}
ds^2 = -\kappa^2r^{2}dt'^{2} + dr^{2}.
\end{align}
Performing a Wick rotation $t' = i\tau$, the Euclidean line element reads
\begin{align}
ds_{E}^{2} = \kappa^{2}r^{2}d\tau^{2} + dr^{2}.
\end{align}
This is precisely the metric of a plane in polar coordinates $(\phi, r)$ with the identification $\phi = \kappa\tau$.  
Regularity at the origin $r=0$ requires the angular coordinate $\phi$ to be periodic with period $2\pi$; hence $\kappa\tau$ must be periodic with period $\kappa\beta = 2\pi$, where $\beta$ is the period in $\tau$. Therefore
\begin{align}
\beta = \frac{2\pi}{\kappa}.
\end{align}
In statistical physics the period $\beta$ of imaginary time is the inverse temperature, as periodicity in imaginary time gives Matsubara frequencies. Consequently, the Hawking temperature associated with the horizon is
\begin{align}
T_{\mathrm{H}} = \frac{1}{\beta} = \frac{\hbar\kappa}{2\pi k_B}.
\end{align}

\section{Extraction of Hawking temperature from density-density correlations}

We consider linear perturbations on top of a stationary background field
$\psi_0(x)$, which can be expanded in Bogoliubov modes as
\begin{equation}
\delta\psi(x,t)
=
\int d\omega
\left[
u_\omega(x)\,\hat b_\omega\,e^{-i\omega t}
+
v_\omega^*(x)\,\hat b_\omega^\dagger\,e^{+i\omega t}
\right].
\label{eq:bogoliubov_expansion}
\end{equation}
The corresponding density fluctuation reads
\begin{equation}
\delta n(x,t)
=
\psi_0^*(x)\,\delta\psi(x,t)
+
\psi_0(x)\,\delta\psi^*(x,t),
\end{equation}
which implies that the density operator couples to the Bogoliubov
combination $u_\omega(x)+v_\omega(x)$. In the presence of an acoustic horizon, the Bogoliubov spectrum contains outgoing Hawking modes outside the horizon and negative energy partner modes inside. For a given frequency $\omega$, the associated annihilation operators $\hat b_H(\omega)$ and $\hat b_P(\omega)$ are related to the incoming modes by a Bogoliubov transformation,
\begin{align}
\hat b_H(\omega)
&=
\alpha_\omega \hat b_+(\omega)
+
\beta_\omega \hat b_-^\dagger(\omega),
\\
\hat b_P(\omega)
&=
\alpha_\omega \hat b_-(\omega)
+
\beta_\omega \hat b_+^\dagger(\omega),
\end{align}
with $|\alpha_\omega|^2-|\beta_\omega|^2=1$, $\langle \hat b_H(\omega)\hat b_P(\omega)\rangle=\alpha_\omega\beta_\omega .$
For the incoming vacuum, the Hawking occupation is
$n_\omega = |\beta_\omega|^2$. Therefore, $|\langle \hat b_H(\omega)\hat b_P(\omega)\rangle|^2=(n_\omega+1)n_\omega$. The equal-time density-density correlation function
\begin{align}
\nonumber
G^{(2)}(x,x')
&=\langle n(x)\rangle\langle n(x^{\prime})\rangle
[g^{(2)}(x,x^{\prime})-1]
\\
&=\langle \delta n(x)\,\delta n(x') \rangle
\end{align}
contains a contribution arising from correlated Hawking partner pairs. Retaining only the cross-horizon terms and using Eq.~\ref{eq:bogoliubov_expansion}, one finds
\begin{align*}
G^{(2)}(x,x')
=
\int d\omega\,
(u_\omega+v_\omega)_H
(u_\omega+v_\omega)_P
\,\\
\times
\langle \hat b_H(\omega)\hat b_P(\omega)\rangle
\,e^{i(k_H x+k_P x')}
+\text{c.c.}
\label{eq:g2_modes}
\end{align*}
where $k_H$ and $k_P$ are the wavevectors of the Hawking and partner modes, respectively, where the conservation of energy enforces frequency matching between Hawking and partner
modes, such that $\omega=\omega_H(k_H)=\omega_P(k_P)$. Because the two modes propagate away from the horizon with different group velocities, the correlations are localized along a straight line in the $(x,x')$ plane corresponding to equal emission times. Projecting $G^{(2)}(x,x')$ along this tangential direction integrates out the emission time and isolates correlations at fixed frequency $\omega$. A subsequent Fourier transform selects the corresponding wavevector pair $(k,k')$,
\begin{align*}
\widetilde G^{(2)}(k,k')
&=
\int dx\,dx'\,
G^{(2)}(x,x')\,e^{-i(kx+k'x')}, \\
&=\int d\omega\,
(u_\omega+v_\omega)_H
(u_\omega+v_\omega)_P\langle \hat b_H(\omega)\hat b_P(\omega)\rangle \\
&\qquad\times \delta(k-k_H)\,\delta(k'-k_P)
+\text{c.c.}
\end{align*}
The extracted quantity is therefore proportional to
\begin{equation}
\big|\widetilde G^{(2)}(k_H,k_P)\big|^2\propto\big|\langle \hat b_H(\omega)\hat b_P(\omega)\rangle\big|^2 = (n_\omega+1)n_\omega,
\label{eq:nnp1}
\end{equation}
which reflects the two-mode squeezing structure of the Hawking process rather than the single particle occupation $n_\omega$. Assuming a thermal Hawking distribution $n_\omega=1/[\exp(\hbar\omega/k_BT_H)-1]$, Eq.~\ref{eq:nnp1} provides a direct determination of the Hawking temperature $T_H$ from the measured correlation spectrum\cite{munoz2019observation}.

\section{SUPPLEMENTARY FIGURES}\label{IV}
\begin{figure*}[h]
\centering
\includegraphics[width=0.7\linewidth]{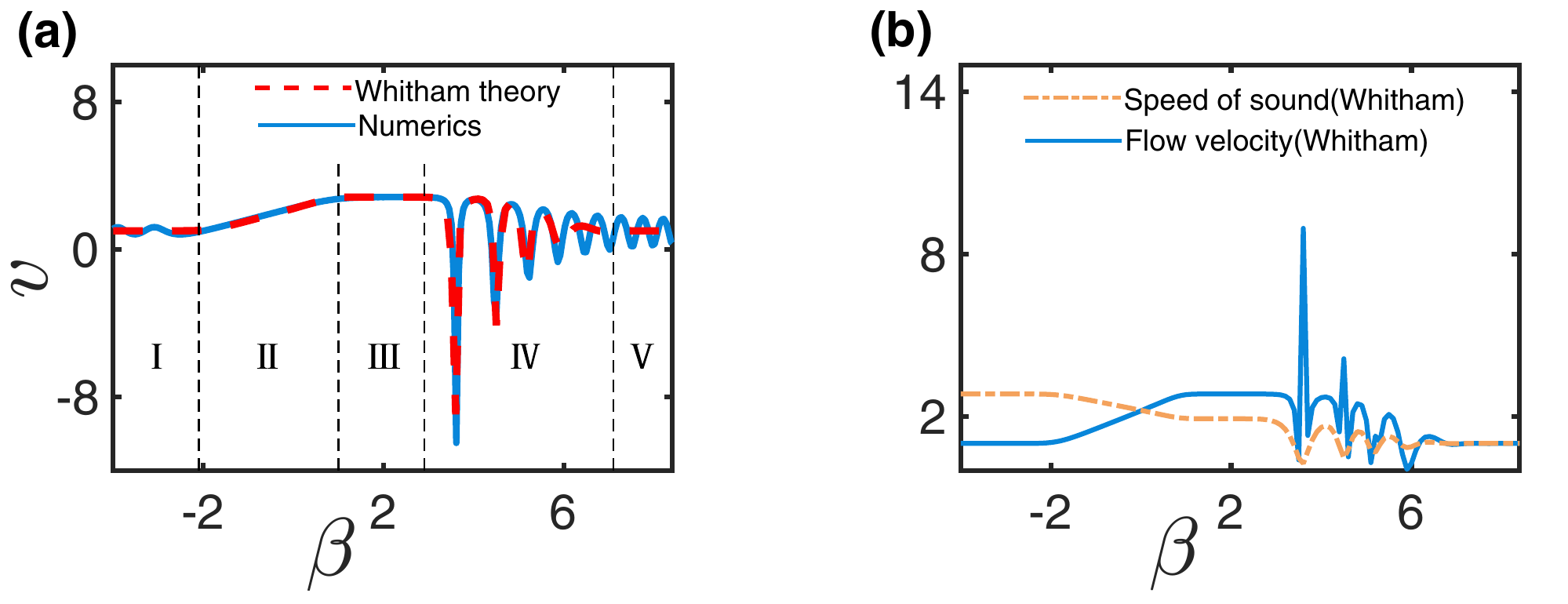}
\caption{\footnotesize (a) Waveform structure of the velocity $v$; (b) Comparison of flow velocity and sound speed based on the Whitham theory.
} \label{figS1}
\end{figure*}
\begin{figure*}[h]
\centering
\includegraphics[width=0.65\linewidth]{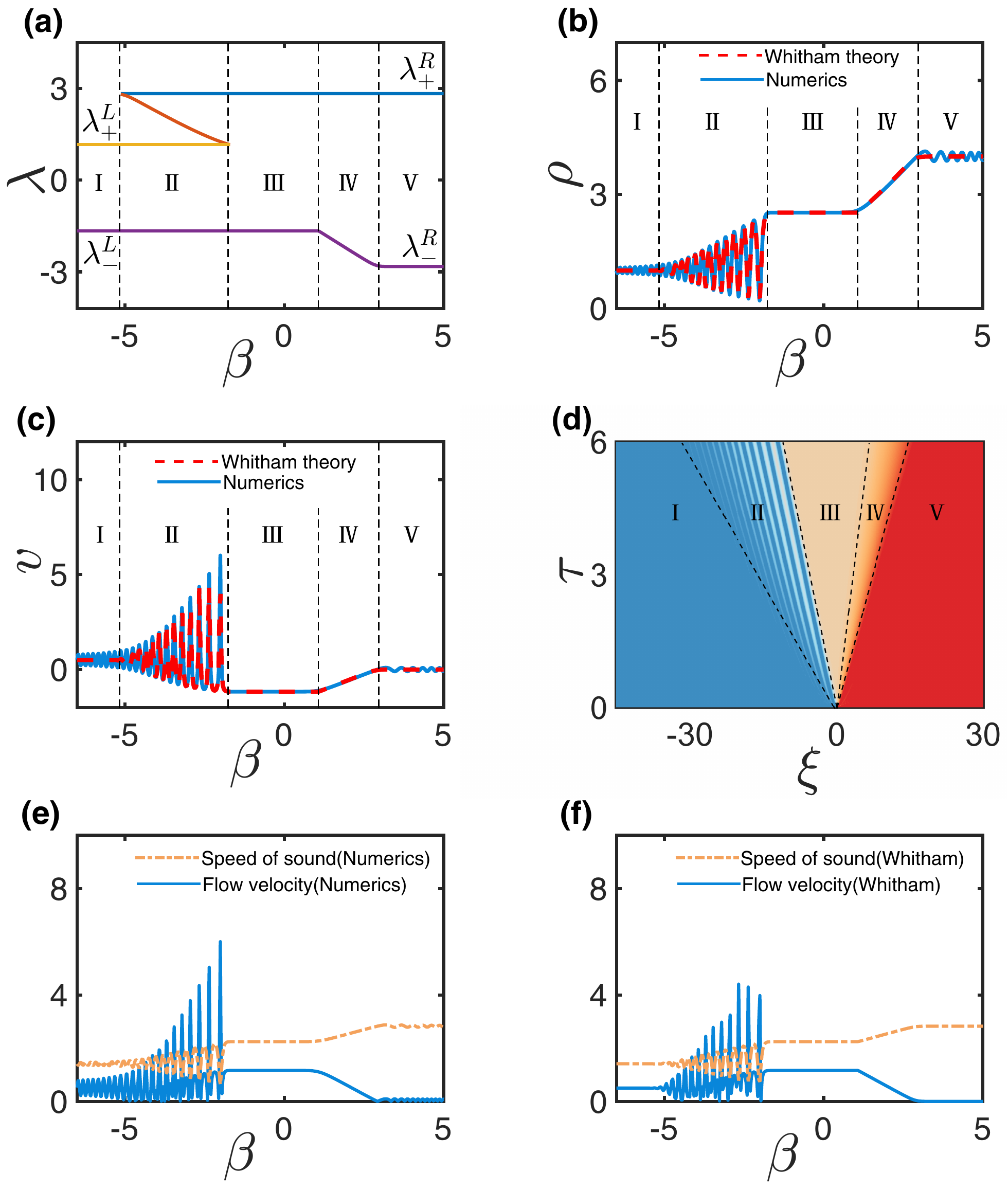}
\caption{\footnotesize (a) Distribution of Riemann invariants; (b, c) Waveform structures of the density $\rho$ and the velocity $v$, respectively; (d) Spatiotemporal evolution of the density $\rho$; (e) Comparison between numerical results of flow velocity and sound speed; (f) Comparison of flow velocity and sound speed based on the Whitham theory. In (b) and (c), the red dashed line represents the result from Whitham theory, and the blue solid line represents the result from numerical computation. In (e) and (f), the orange dash-dotted line represents the sound speed, and the blue solid line represents the flow velocity. The parameters are: $\alpha =2, \tau = 6, (\rho_R, v_R)=(4,0), (\rho_0, v_0)=(1,0.5)$.
} \label{fig2}
\end{figure*}
\begin{figure*}[h]
\centering
\includegraphics[width=0.65\linewidth]{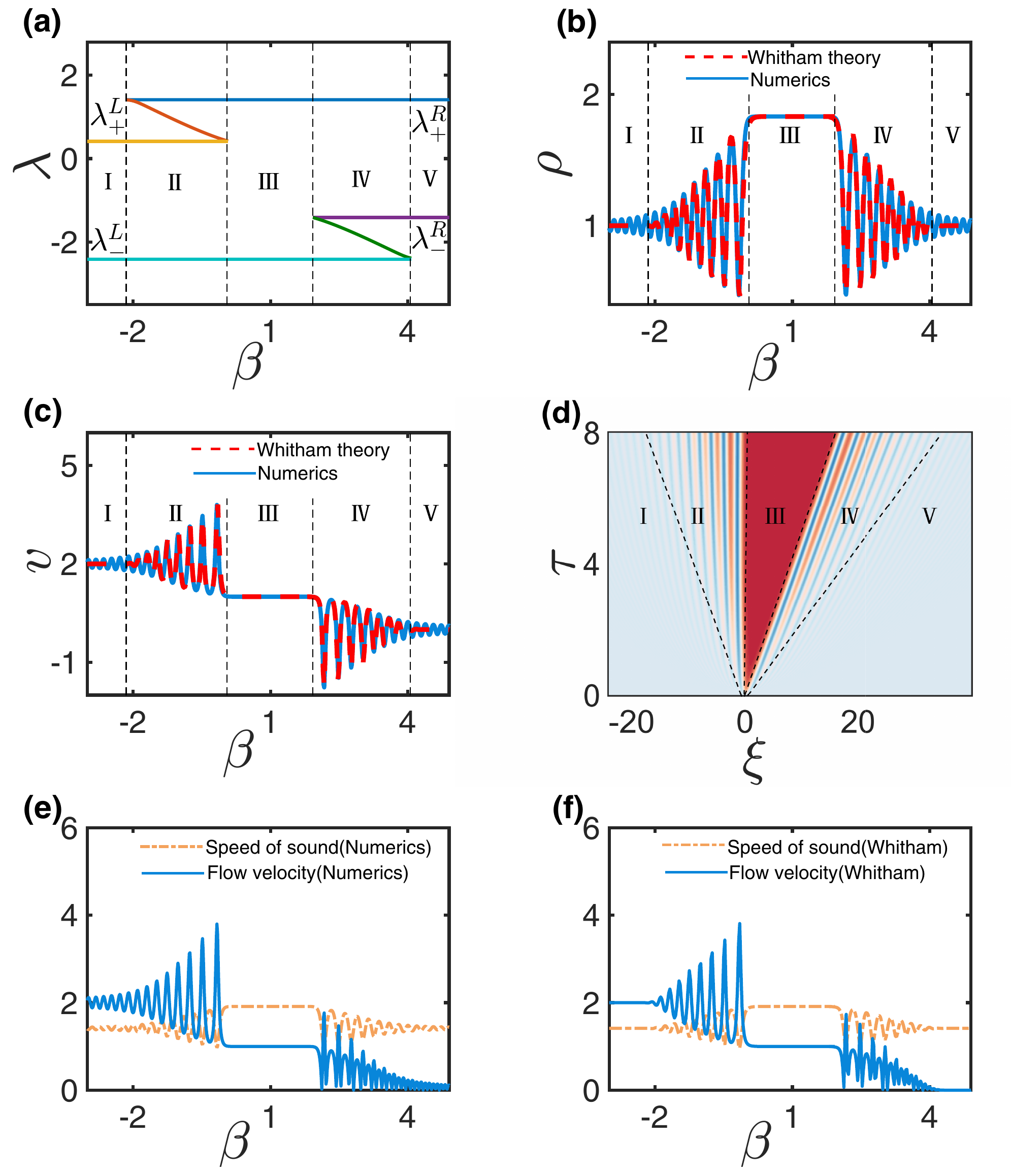}
\caption{\footnotesize (a) Distribution of Riemann invariants; (b, c) Waveform structures of the density $\rho$ and the velocity $v$, respectively; (d) Spatiotemporal evolution of the density $\rho$; (e) Comparison between numerical results of flow velocity and sound speed; (f) Comparison of flow velocity and sound speed based on the Whitham theory. In (b) and (c), the red dashed line represents the result from Whitham theory, and the blue solid line represents the result from numerical computation. In (e) and (f), the orange dash-dotted line represents the sound speed, and the blue solid line represents the flow velocity.  The parameters are: $\alpha =2, \tau = 8, (\rho_R, v_R)=(1,0), (\rho_0, v_0)=(1,2)$.
} \label{fig4}
\end{figure*}
\begin{figure*}[h]
\centering
\includegraphics[width=0.8\linewidth]{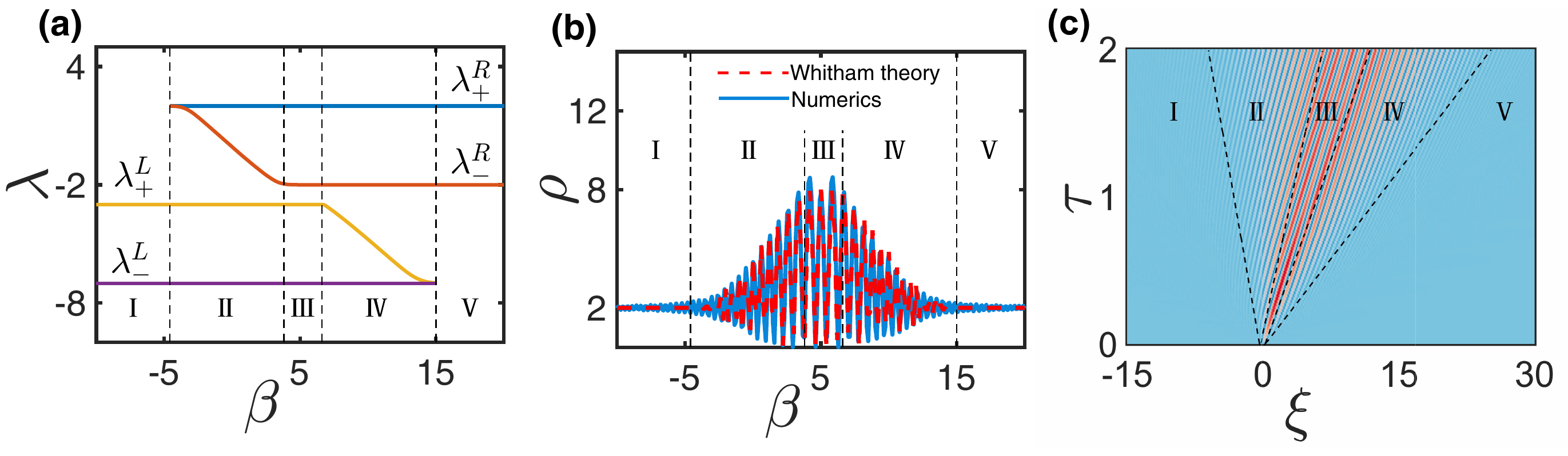}
\caption{\footnotesize (a) Distribution of Riemann invariants; (b) Waveform structures of the density $\rho$; (c) Spatiotemporal evolution of the density $\rho$ up to $\tau=2$; In (b), the red dashed line represents the result from Whitham theory, and the blue solid line represents the result from numerical computation. (a) and (b) present the results at $\tau = 1.5$.The parameters are: $\alpha =2, (\rho_R, v_R)=(2,0), (\rho_0, v_0)=(2,10)$.
} \label{fig5}
\end{figure*}
\begin{figure*}[h]
\centering
\includegraphics[width=0.65\linewidth]{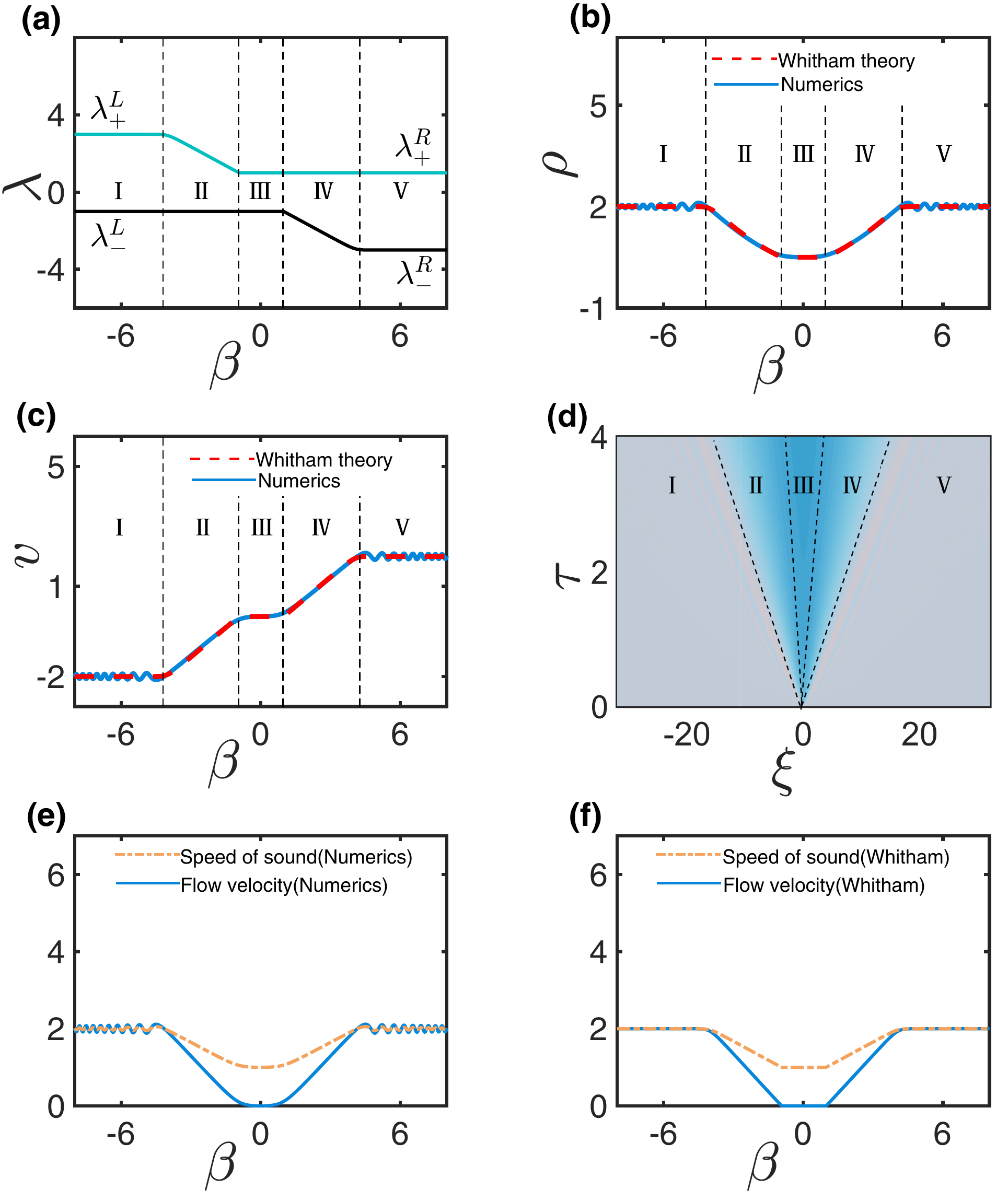}
\caption{\footnotesize (a) Distribution of Riemann invariants; (b, c) Waveform structures of the density $\rho$ and the velocity $v$, respectively; (d) Spatiotemporal evolution of the density $\rho$; (e) Comparison between numerical results of flow velocity and sound speed; (f) Comparison of flow velocity and sound speed based on the Whitham theory. In (b) and (c), the red dashed line represents the result from Whitham theory, and the blue solid line represents the result from numerical computation. In (e) and (f), the orange dash-dotted line represents the sound speed, and the blue solid line represents the flow velocity. The parameters are: $\alpha=2, \tau = 4, (\rho_R, v_R)=(2,2), (\rho_0, v_0)=(2,-2)$.
} \label{fig6}
\end{figure*}
\begin{figure*}[h]
\centering
\includegraphics[width=0.65\linewidth]{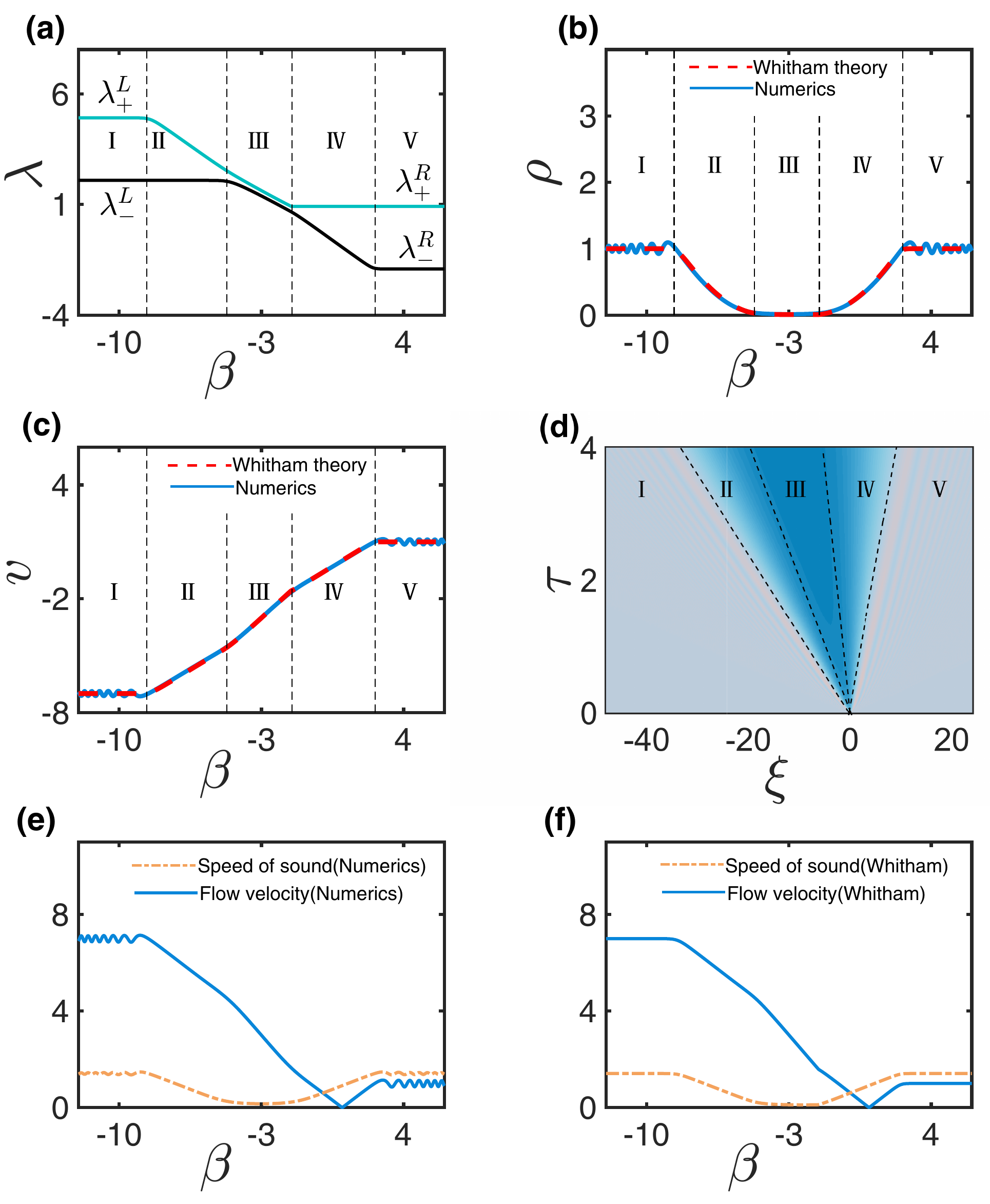}
\caption{\footnotesize (a) Distribution of Riemann invariants; (b, c) Waveform structures of the density $\rho$ and the velocity $v$, respectively; (d) Spatiotemporal evolution of the density $\rho$; (e) Comparison between numerical results of flow velocity and sound speed; (f) Comparison of flow velocity and sound speed based on the Whitham theory. In (b) and (c), the red dashed line represents the result from Whitham theory, and the blue solid line represents the result from numerical computation. In (e) and (f), the orange dash-dotted line represents the sound speed, and the blue solid line represents the flow velocity. The parameters are: $\alpha=2, \tau = 4, (\rho_R, v_R)=(1,1), (\rho_0, v_0)=(1,-7)$.
} \label{fig7}
\end{figure*}
\begin{figure*}[h]
\centering
\includegraphics[width=0.8\linewidth]{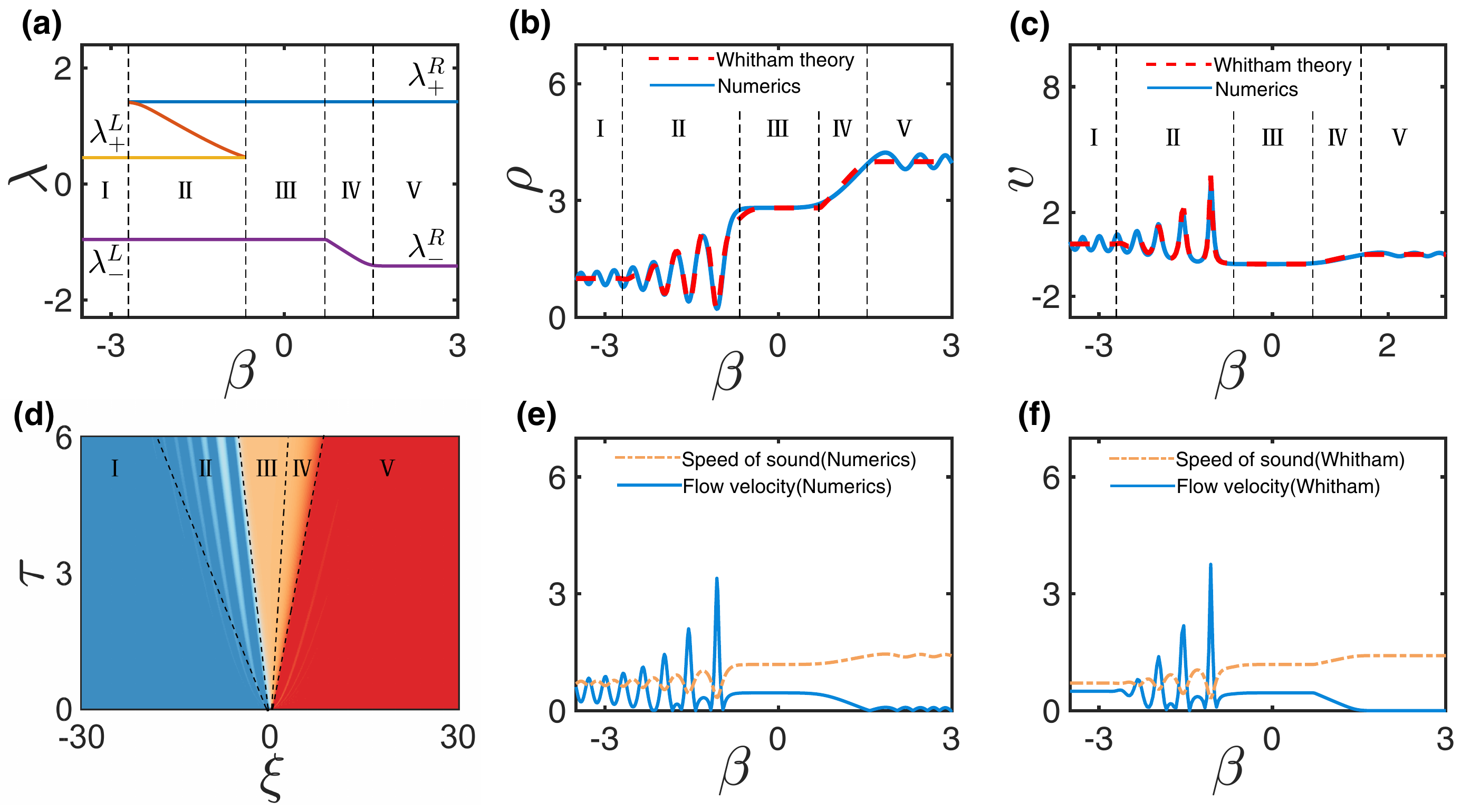}
\caption{\footnotesize (a) Distribution of Riemann invariants; (b, c) Waveform structures of the density $\rho$ and the velocity $v$, respectively; (d) Spatiotemporal evolution of the density $\rho$; (e) Comparison between numerical results of flow velocity and sound speed; (f) Comparison of flow velocity and sound speed based on the Whitham theory. In (b) and (c), the red dashed line represents the result from Whitham theory, and the blue solid line represents the result from numerical computation. In (e) and (f), the orange dash-dotted line represents the sound speed, and the blue solid line represents the flow velocity. The parameters are: $\alpha=0.5, \tau = 6, (\rho_R, v_R)=(4,0), (\rho_0, v_0)=(1,0.5)$.
} \label{fig8}
\end{figure*}
\begin{figure*}[h]
\centering
\includegraphics[width=0.8\linewidth]{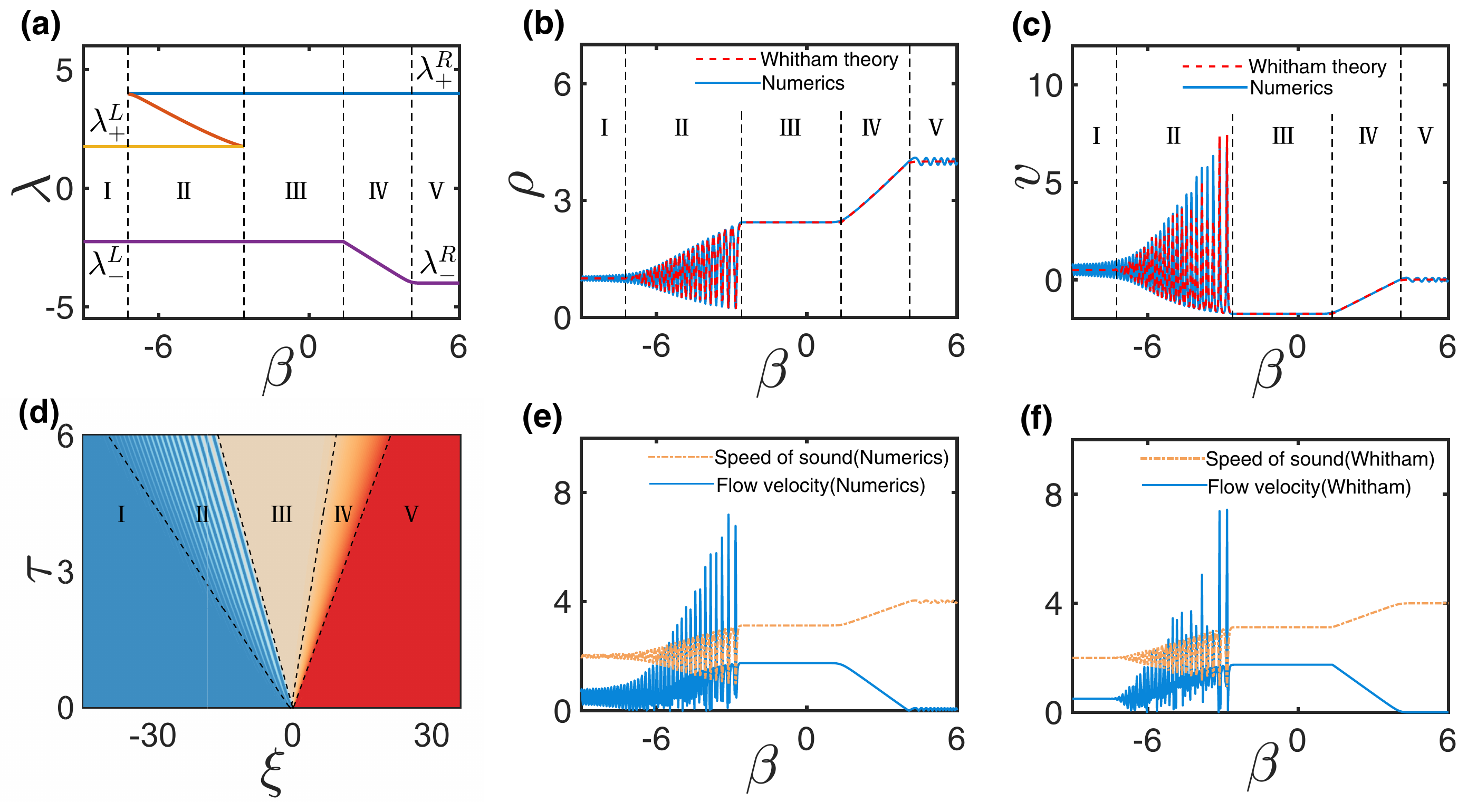}
\caption{\footnotesize (a) Distribution of Riemann invariants; (b, c) Waveform structures of the density $\rho$ and the velocity $v$, respectively; (d) Spatiotemporal evolution of the density $\rho$; (e) Comparison between numerical results of flow velocity and sound speed; (f) Comparison of flow velocity and sound speed based on the Whitham theory. In (b) and (c), the red dashed line represents the result from Whitham theory, and the blue solid line represents the result from numerical computation. In (e) and (f), the orange dash-dotted line represents the sound speed, and the blue solid line represents the flow velocity. The parameters are: $\alpha =4, \tau = 6, (\rho_R, v_R)=(4,0), (\rho_0, v_0)=(1,0.5)$.
} \label{fig9}
\end{figure*}

\begin{figure*}[h]
    \centering
    \includegraphics[width=0.9\linewidth]{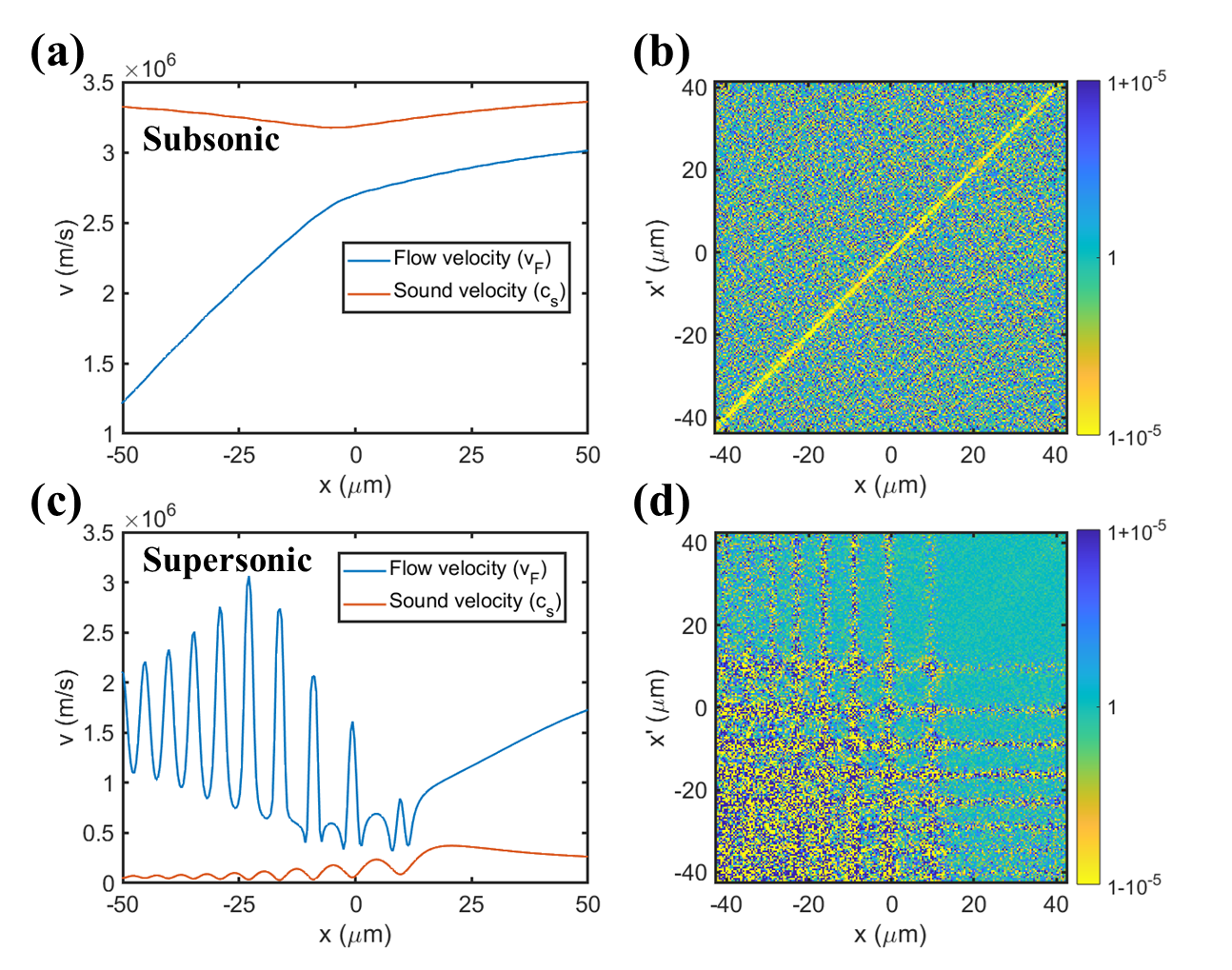}
    \caption{(a) Subsonic velocity field where $v_F<c_s$; (b) $g^{(2)}(x,x^\prime)$ for the subsonic regime; (c) Supersonic velocity field where $v_F>c_s$; (d) $g^{(2)}(x,x^\prime)$ for the subsonic regime. The number of samplings of initial quantum fluctuation is $10^4$. In both cases no acoustic blackhole is detected from the second-order correlation diagram.}
    \label{fig:si_g2}
\end{figure*}


\bibliography{apssamp}